\documentclass[conference]{IEEEtran}
\IEEEoverridecommandlockouts
\usepackage{cite}
\usepackage{amsmath,amssymb,amsfonts}
\usepackage{algorithmic}
\usepackage{graphicx}
\usepackage{textcomp}
\usepackage{xcolor}

\def\BibTeX{{\rm B\kern-.05em{\sc i\kern-.025em b}\kern-.08em
		T\kern-.1667em\lower.7ex\hbox{E}\kern-.125emX}}

\usepackage[a4paper,
bindingoffset=0.2in,
left=1.32cm,
right=1.32cm,
top=0.78in,
bottom=4.3cm,
]{geometry}

\usepackage{tipa}
\usepackage{bm}
\usepackage{bbm}
\usepackage{mathrsfs}
\usepackage{amsfonts}
\usepackage{cite,epsfig,graphicx}
\usepackage{amssymb,amsmath}
\usepackage{indentfirst}
\usepackage{subfig}
\usepackage{epstopdf}
\usepackage{lipsum}
\usepackage{tikz} 
\usetikzlibrary{matrix,shapes,arrows,positioning,chains, calc, backgrounds}
\usepackage{multirow}
\usepackage{colortbl}
\usepackage[hyphens]{url}
\usepackage{multirow}
\usepackage{tablefootnote}
\usepackage{booktabs}

\begin{document}
\title{Cryptocurrency Giveaway Scam with \\ YouTube Live Stream }

\makeatletter
\newcommand{\linebreakand}{%
\end{@IEEEauthorhalign}
\hfill\mbox{}\par
\mbox{}\hfill\begin{@IEEEauthorhalign}
}
\makeatother

\author{
	\IEEEauthorblockN{Iman Vakilinia \\ 
	\IEEEauthorblockA{ School of Computing, \\ University of North Florida, \\ Jacksonville, FL, USA, \\ i.vakilinia@unf.edu}}
}

\maketitle

\begin{abstract}

This paper investigates the cryptocurrency giveaway scam with the YouTube live stream. In this scam scheme, the scammer streams a recorded video of a famous person in a YouTube live stream, and annotates the video with a cryptocurrency giveaway announcement. In the announcement, the victims are encouraged to visit the scammer's webpage for a cryptocurrency giveaway. 
The webpage is designed intelligently to deceive victims. The scam claims that whatever donation the victim sends to a cryptocurrency wallet address, the giveaway scheme will double the donated amount and immediately send it back to the victim.
By analyzing the scammers' wallet addresses, it can be seen that scammers could steal a significant amount of money in a short time. 
After analyzing the scammers' techniques, this paper discusses the countermeasures that can be applied to mitigate such a fraudulent activity in the future.

\end{abstract}


\section{Introduction}\label{sec:intro}

Recently, scammers are actively utilizing YouTube live streams for cryptocurrency giveaway scams.
The scam scheme uses the recorded videos of tech celebrities, such as Elon Musk, Jack Dorsey, Bill Gates, Vitalik Buterin, and Michael Saylor to promote a fake cryptocurrency giveaway. The video is annotated with an announcement to direct victims to the scammer's webpage for more details about the giveaway. On the webpage, the scammer shares a cryptocurrency wallet address and requests a donation claiming that the giveaway will double the donated amount and immediately send it back to the victim.

According to a report by cybersecurity firm GROUP-IB~\cite{group}, between February 16 and 18, 2022, the scammers ran 36 fabricated cryptocurrency giveaway YouTube streams that attracted more than 165,000 viewers. Group-IB’s estimated that the wallets controlled by the scammers received more than \$1.6 million in 281 transactions.
Based on the report, all these channels have supposedly been either hacked or purchased on the underground market.

According to the Tenable report~\cite{tenable2}, scammers have stolen over \$9 million dollars using YouTube live stream during the three days of May 7th to May 9th of 2021. In another research study, scammers have stolen \$8.9 million dollars over one month period from October/21/2021 to November/21/2021~\cite{tenable}. In the former study, the scammers have used footage of the notable persons including ``Michael Saylor, chairman, and CEO of MicroStrategy and a fervent supporter of Bitcoin; Vitalik Buterin, Ethereum co-founder; Charles Hoskinson, Cardano founder, and Ethereum co-founder; Brad Garlinghouse, CEO of Ripple Labs; Elon Musk, CEO of Tesla and SpaceX".

Several lawsuits have been filed regarding this issue against YouTube.
Apple co-founder Steve Wozniak filed a lawsuit against YouTube over videos that used his image to promote a cryptocurrency giveaway in 2020. This lawsuit has been rejected as a California state judge said YouTube and its parent company, Google LLC, are protected by federal law from responsibility for content posted by users.

Wozniak argued in his lawsuit that Section 230 of the Communications Decency Act should not apply because YouTube not only failed to remove the fraudulent videos but ``materially contributed" to the scam by selling targeted ads driving traffic to the videos and falsely verifying the YouTube channels that carried the videos.
However, Santa Clara County Superior Court Judge said those factors were not enough to overcome the immunity provided by Section 230~\cite{wozni}. 

Ripple and its CEO Brad Garlinghouse have also filed a lawsuit against YouTube in 2020 claiming that the platform knowingly profits from the actions of the scammers, despite having the ability to stop them~\cite{ripple}. Later in 2021, Ripple ended this lawsuit expressing that the companies have reached a resolution to work together to prevent, detect and take down these scams. However, details regarding the nature of the settlement between Ripple and YouTube remained confidential~\cite{ripple22}.

\begin{figure*}[ht]
	\centering
	\includegraphics[width=0.9\textwidth]{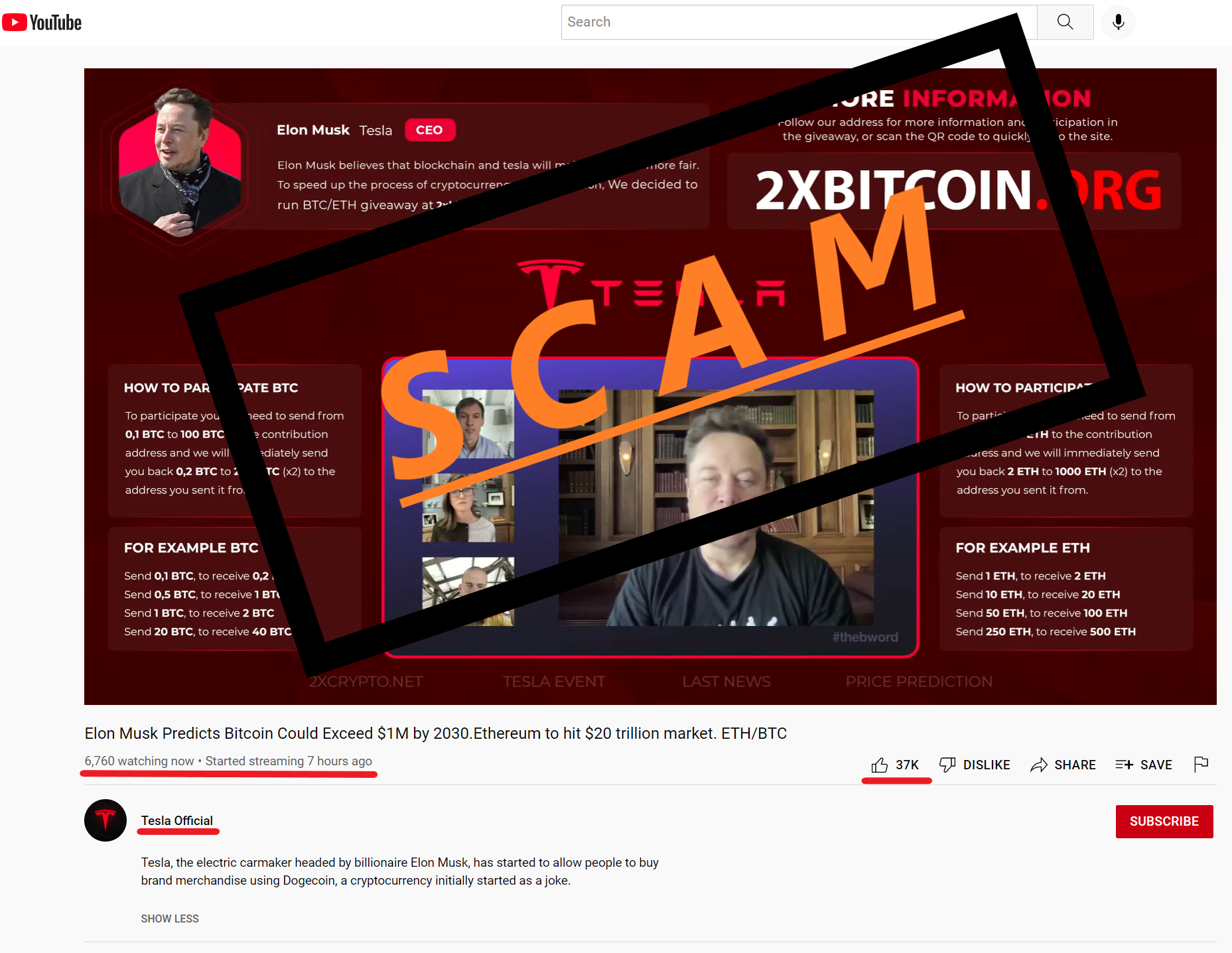}
	\caption{A screenshot of a YouTube Live stream giveaway scam}
	\label{fig:1}
\end{figure*}

On the other hand, Google has announced a Phishing campaign that targets YouTube creators with cookie theft malware on October 20, 2021~\cite{goog}. According to this report, a large number of hijacked channels were rebranded for cryptocurrency scam live-streaming. The channel name, profile picture, and content were all replaced with cryptocurrency branding to impersonate large tech or cryptocurrency exchange firms. The attacker live-streamed videos promising cryptocurrency giveaways in exchange for an initial contribution.
On account-trading markets, hijacked channels ranged from \$3 USD to \$4,000 USD depending on the number of subscribers.

Following the history of active and successful cryptocurrency giveaway scams with YouTube live streams, this paper investigates the current status of this scam. To this end, this paper has investigated the YouTube live streams cryptocurrency giveaway scams which have been carried out on 5/15/2022 and 5/16/2022. 
Once a live stream scam was found, the webpage advertised in the video was visited to analyze the scam in more detail.
The result shows that this scam is still effective and actively ongoing.

\section{YouTube Giveaway Scam}

\begin{figure}[h]
	\centering
	\includegraphics[width=0.45\textwidth]{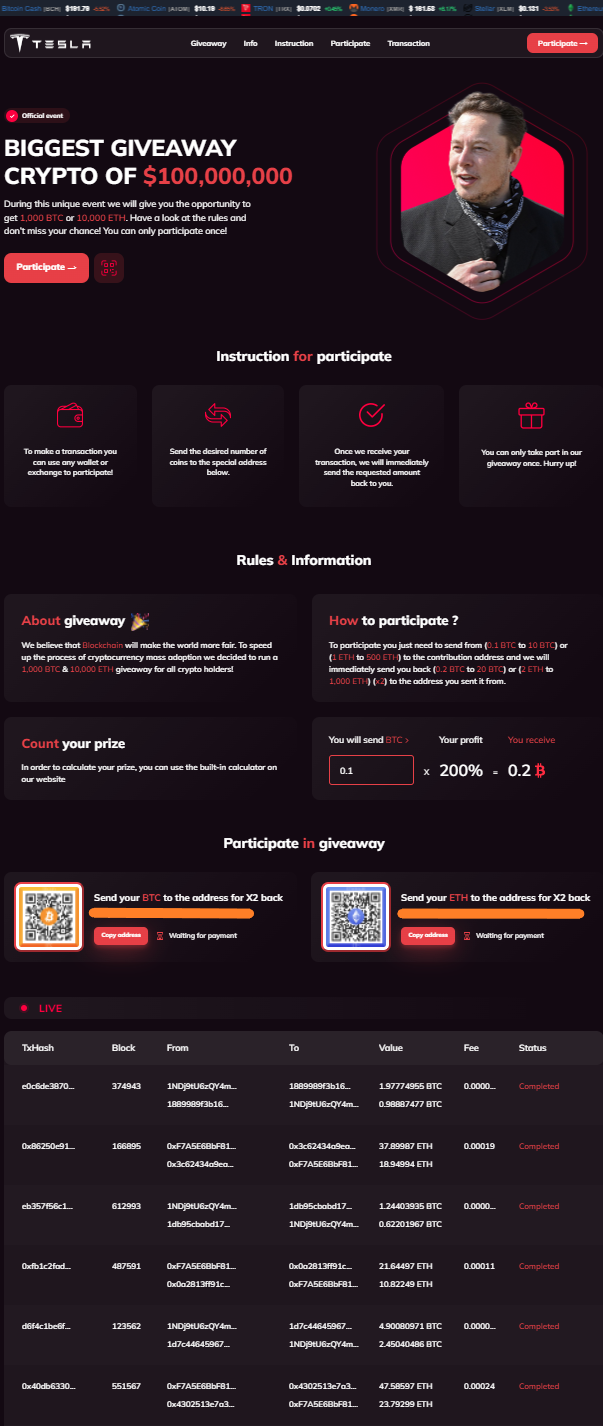}
	\caption{Cryptocurrency scam webpage}
	\label{fig:2}
\end{figure}

A combination of ``bitcoin, ethereum, live, giveaway, musk" keywords have been used to find the cryptocurrency giveaway live stream scams on YouTube.
In total 18 scam live streams have been found. Out of these videos, 13 were streaming an interview of Tesla CEO Elon Musk, Twitter CEO Jack Dorsey, and Ark Invest CEO Cathie Wood which was originally recorded in July 2021~\cite{interv}. Two videos were streaming ``Third Row Tesla Podcast – Episode 7 - Elon Musk's Story"\cite{interv2} which was originally recorded in Feb 2020. The remaining three videos were streaming Michael Saylor's interview which was originally recorded in November 2021~\cite{interv3}.

For all cases, the scammer has changed the name of the account to one of the following names: ``Tesla Official, Tesla Live, Tesla, Tesla NEWS, MicroStrategy". In most cases, the previous shared contents of the YouTube account have been all removed. However, in some cases there were some crypto unrelated videos existed in the channel. By checking the ``Store" and ``community" tabs, it could be seen that in some cases, there are old crypto unrelated posts which have not been removed. By checking the about section, it could be seen that all of the accounts have been created before 2015. 
Observing these facts, it is reasonable to assume that these accounts have been hacked. Moreover,, a YouTuber with a large number of subscribers would not normally risk loosing its account for running a scam. 

The number of scam live stream viewers were varied from 900 to 46,000. Most of the live streams had around (3K - 10K) likes. However, one scam live stream which was streaming ``Third Row Tesla Podcast – Episode 7" had ``104K" likes. This live stream had 16,456 watchers at the time of checking.  Surprisingly, the original video which has published on Feb 2020~\cite{interv3} has 27K likes. 
It is questionable if the scammers have applied/purchased fake likes or all likes are from viewers. 

By checking the YouTube account, it could be seen that the scammer streams multiple live streams at the moment as can be seen in figure~\ref{fig:3}. 
It seems that, this strategy can promote the live streams. The live streams are recommended to viewers in the suggested video panel, and seeing multiple live streams of a same event can deceive viewers that the event is genuine as several channels are covering it live.

\begin{figure}[ht]
	\centering
	\includegraphics[width=0.45\textwidth]{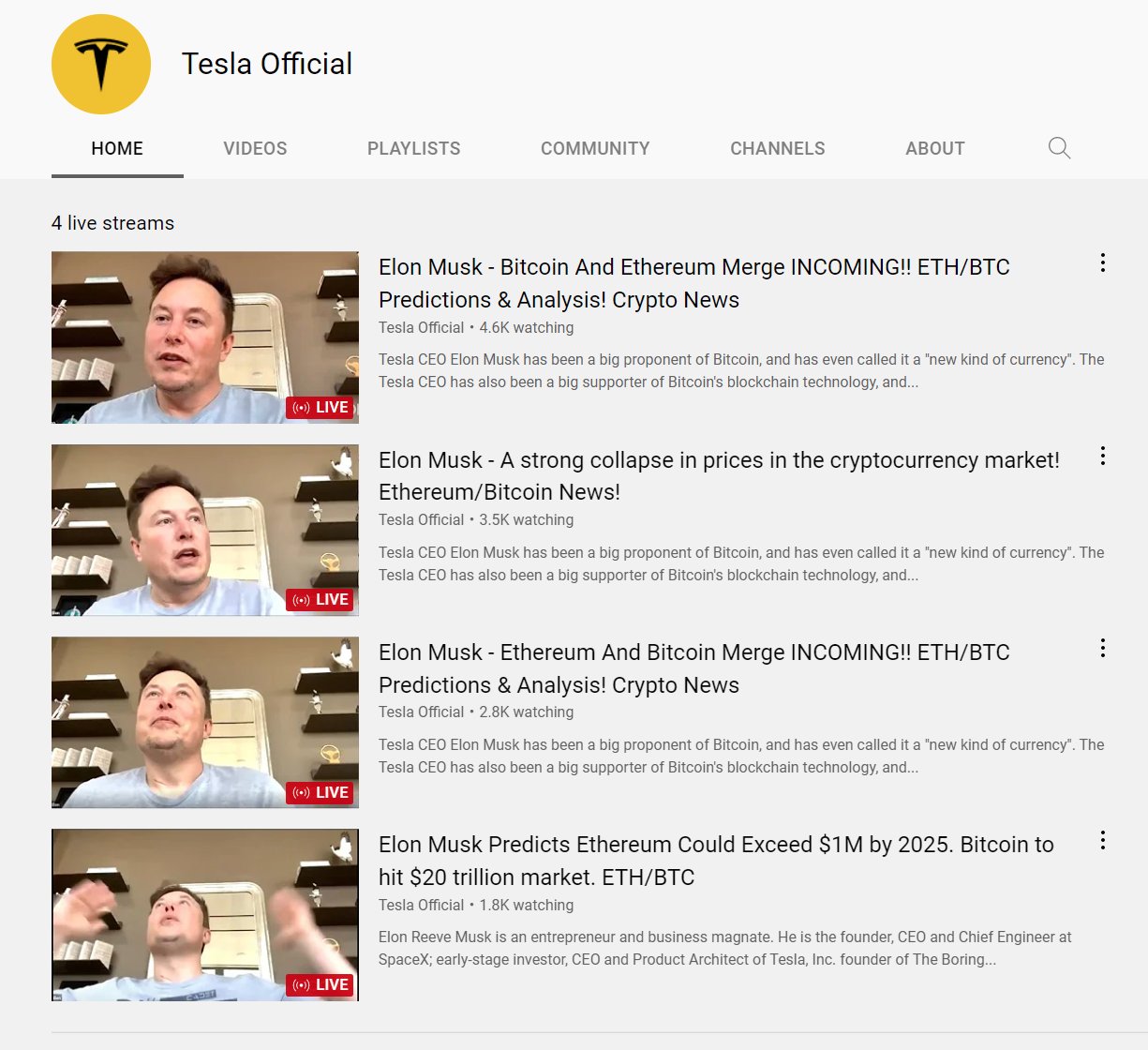}
	\caption{Multiple giveaway scam live streams from one account}
	\label{fig:3}
\end{figure}

\subsection{Giveaway Webpage Analysis}

One of the giveaway webpage is shown in fig~\ref{fig:2}. Other giveaway webpages have used very similar platforms. To make the website looks legitimate, scammers have designed a good looking website with several sections as follows:

\vspace{1mm}
\textbf{Live Cryptocurrency prices widget}

The website displays the live price of cryptocurrncies on top of the website. To this end, the scammer has used the coinlib widget embedded code~\cite{widget}. 

\vspace{1mm}
\textbf{Live Chat support}

The website is equipped with a live chat support. To this end, the website has used ``tawk.to" chat service.
To verify if the chat support is a programmed bot or a human, the author has communicated with the chat support. By asking different questions, it was perceived that a human is replying to the messages. 

\vspace{1mm}
\textbf{Fake live transactions}

The webpage demonstrates a set of fake live transactions at the bottom. New fake transactions are added periodically in a short time.  

\vspace{1mm}
\textbf{Deceptive Texts}

The following deceptive texts have been used to precipitate victims.

\begin{itemize}
	\item You can only do it one time!
	\item Giveaway will be over soon!  
	\item The fund is limited!
\end{itemize}

In some cases, a diagram was showing the decrease of available funds.

\subsection{Domain names registration information}

Table~\ref{dns} demonstrates the scams' domain names registration information received from whois.com.

\begin{table*}[h]
	\caption{Domain name registration information received from whois.com}
	\begin{center}
		\begin{tabular}{ |m{1.8cm}|m{3.2cm}|m{1.5cm}|m{3cm}|m{2cm}|m{2.5cm}|} 
			\hline
			\rowcolor{lightgray}  Domain name & Registrar & Registered On & Name Server & Organization/Name & Address 
			\\ [1ex]
			\hline
			teslaget.org & Registrar of Domain Names REG.RU LLC & 2022-05-14 & jade.ns.cloudflare.com wilson.ns.cloudflare.com & SergeiIUdakov & State:Moskovskaia obl/ Country:RU\\
			\hline
			btc-elon.net & Registrar of Domain Names REG.RU LLC& 2022-05-01 & benedict.ns.cloudflare.com sasha.ns.cloudflare.com & Sergei IUdakov & State:Moskovskaia obl/ Country:RU \\
			\hline
			microstrategy.vin & Registrar of Domain Names REG.RU LLC & 2022-05-14 & jade.ns.cloudflare.com wilson.ns.cloudflare.com & SergeiIUdakov & State:Moskovskaia obl/ Country:RU \\
			\hline
			2xbitcoin.org & Registrar of Domain Names REG.RU LLC & 2022-05-15 & raegan.ns.cloudflare.com odin.ns.cloudflare.com & SergeiIUdakov & State:Moskovskaia obl/ Country:RU \\
			\hline
			fundtesla.us & PDR Ltd. d/b/a PublicDomainRegistry.com & 2022-05-15 & ns2.ddos-guard.net ns1.ddos-guard.net & Alex Jordon & State:Rlga/ Country:LV \\
			\hline
			2xtesla.us & PDR Ltd. d/b/a PublicDomainRegistry.com & 2022-05-15 & ns2.ddos-guard.net ns1.ddos-guard.net & Alex Jordon & State:Rlga/	Country:LV \\
			\hline
			twitup.org & Registrar of Domain Names REG.RU LLC & 2022-05-11 & pat.ns.cloudflare.com \newline maciej.ns.cloudflare.com & Private Person & State:Moscow/ \newline Country:DE \\
			\hline
			ether2x.net & Registrar of Domain Names REG.RU LLC & 2022-05-12 & ns1.hosting.reg.ru	ns2.hosting.reg.ru & Ivan Ivanov & State:Moscow/ \newline Country:RU \\
			\hline
			ms-btc.us & Sav.com LLC & 2022-05-10 & aspen.ns.cloudflare.com	clayton.ns.cloudflare.com & lewal52205 lewal52205 & State:Ms/ \newline Country:RU \\
			\hline
			tesla-drops.org & Registrar of Domain Names REG.RU LLC & 2022-05-12 & ns1.shelter.to ns2.shelter.to & Wild Boar Shelter & State:Omsk Region/ \newline Country:RU \\
			\hline
			ethtesla.us & PDR Ltd. d/b/a PublicDomainRegistry.com & 2022-05-13 & anuj.ns.cloudflare.com 			sierra.ns.cloudflare.com & Muni Muni & State:Helsinki/ \newline Country:FI \\
			\hline
			elontesla2x.com &NICENIC INTERNATIONAL GROUP CO., LIMITED & 2022-05-09 & ns1.my-ndns.com \newline ns2.my-ndns.com & DOMAIN PRIVACY & State:HK/ Country:CN \\
			\hline
			elon-fund.net & Hosting Concepts B.V. d/b/a Registrar.eu & 2022-05-04 & byron.ns.cloudflare.com lisa.ns.cloudflare.com & Whois Privacy Protection Foundation & State:Zuid-Holland/ Country:NL \\
			\hline
			ms2x.one & PDR Ltd. d/b/a PublicDomainRegistry.com & 2022-05-09 & clara.ns.cloudflare.com houston.ns.cloudflare.com & N/A & State:Kiev/ Country:UA \\
			\hline
			muskelon.io & NetEarth One Inc. dba NetEarth & 2022-05-11 & clara.ns.cloudflare.com 			houston.ns.cloudflare.com & N/A & State:Kiev/ Country:UA \\
			\hline
		\end{tabular}
		\label{dns}
	\end{center}
\end{table*}

Most webpages have used cloudflare Content Delivery Network service as it can be seen in the nameserver field.
It is worth mentioning that by checking these websites in the next days, the Cloudflare has labeled the websites as unsafe showing a message of ``\textit{Warning: Suspected Phishing Site Ahead! This link has been flagged as phishing. We suggest you avoid it.}"

Domain names have been registered a few days before the giveaway scam. Some of the registration information are same for different domain names implying a same campaign behind the giveaway scam.

\subsection{Transactions}

The transaction history of scammers' ethereum and bitcoin addresses are demonstrated in table~\ref{trs} which are taken from etherscan.io and blockchain.com.

\begin{table*}[h]
	\caption{Transaction history received from etherscan.io and blockchain.com}
	\begin{center}
		\begin{tabular}{ |m{1.8cm}|m{6.3cm}|c|c|} 
			\hline
			\rowcolor{lightgray}  Domain name & Eth/Bitcoin addresses & Number of In TXs & Total In \\ [1ex]
			\hline
			\multirow{ 2}{*}{teslaget.org } & 0x2Ed3da051F8a5F8fAA8D6A09c842388d04456Ca4 & 8 & 6.572415868541148 \\
			& bc1q7k92epzh3fxvhz8glr7xckx23p5sw530gse67j & 2 & 
			0.00061852 \\ \hline
			
			\multirow{ 2}{*}{btc-elon.net } & 0x917b3d76B237B59Cb6d8ad7058BC2DFddB768FC3 & 5 & 2.09849628\\
			& 1BFwsZuFHkmtfMF7d13bRVYt8WtvUVRd9t & 10 & 
			0.59769947 \\ \hline
			
			\multirow{ 2}{*}{microstrategy.vin } & 0xE2A9B9f5f14Cb78B8Fc1Fe845c7939036e9f07d5 & 21 & 44.891283571251568\\
			& 1PAJQG4BxuNNwKSS6EsuS4xxVpZ8CbKSJK & 24 & 5.88351915 \\ \hline
			
			\multirow{ 3}{*}{2xbitcoin.org } & 0x9Ca287De08115215A65C6444Ab9275eBb6119aB7 & 23 & 31.607592727646691398\\
			& 0xee1B39a9A1E6A87C24c4FF31b645c58c076106d8 & 24 & 26.972458942987742 \\
			& 1K1YjyjTusN2ijQwQqkYp6HNnoZcdFWJid & 32 & 3.63523266 \\ \hline
			
			\multirow{ 2}{*}{fundtesla.us} & 0xF7A5E6BbF816Be2221ce149bcC5e9CC777A3B7e2 & 19 & 14.662921377974322 \\
			& bc1q7k92epzh3fxvhz8glr7xckx23p5sw530gse67j & 11 & 
			1.31218138 \\ \hline
			
			\multirow{ 2}{*}{2xtesla.us } & 0xF7A5E6BbF816Be2221ce149bcC5e9CC777A3B7e2 & 19 & 14.662921377974322 \\
			& bc1q7k92epzh3fxvhz8glr7xckx23p5sw530gse67j & 11 & 
			1.31218138 \\ \hline
			
			\multirow{ 2}{*}{twitup.org } & 0x4e2bd4a18dA82D14d5764F5c25ad9adef30E86C5 & 14 & 14.789810876541761185\\
			& 1H8eygZVfWpgzwk7yMDXsQcGgYh8cW5dXx & 5 & 0.03516008 \\ \hline

			\multirow{ 2}{*}{ether2x.net } & 0x2Ed3da051F8a5F8fAA8D6A09c842388d04456Ca4 & 8 & 6.572415868541148 \\
			& bc1q7k92epzh3fxvhz8glr7xckx23p5sw530gse67j & 2 & 
			0.00061852 \\ \hline

			\multirow{ 2}{*}{ms-btc.us } & 0x2Ed3da051F8a5F8fAA8D6A09c842388d04456Ca4 & 1 & 0.099527637712372 \\
			& bc1qudd0d3802ja9c806necjxt2v55d7dw2w3hq29a & 1 & 0.25000000 \\ \hline

			\multirow{ 2}{*}{tesla-drops.org } & 0x46a36ff5aE210215348a7fb60B49676749579fE4 & 11 & 20.586943184252219 \\
			& 19Q7qX6DENhKsfXLPcjR6XULPoec5RLEU7 & 14 & 0.44607352 \\ \hline

			\multirow{ 2}{*}{ethtesla.us } & 0x0024a0BEB06e0C9136A4aD96ecEE2D35Aa61F048 & 0 & 0\\
			& 1CJYt8zcnXxTRtfGnEhnuVtnmxTpWL1HEX & 0 & 0 \\ \hline

			\multirow{ 2}{*}{elontesla2x.com } & 0xe9781C03dcA555212a4FB2ff5EC48ff9D6B81Cd0 & 5 & 4.33331649\\
			& 1KCHtxvpJnkkp7wkxTTLvaw6VGRv4wccXR & 8 & 
			1.45012020 \\ \hline

			\multirow{ 2}{*}{elon-fund.net } & 0x3ae7A923A1B318454D69Ef9d902d4A1667d71655 & 16 & 11.882348341814229\\
			& 1P3xLkrzGhtDQZL2do26hnAwXufNc6nguv & 1 & 
			0.50402378  \\ \hline

			\multirow{ 2}{*}{ms2x.one  } & 0x6daCDE98C5FB6F0A3eE4E71c8681216E0C17231B & 4 & 0.6976665645274818 \\
			& 1PY5y459ErUiexiwhpWeCLRcBKfqtxagQk & 2 & 
			0.00209249 \\ \hline

			\multirow{ 2}{*}{muskelon.io } & 0x6daCDE98C5FB6F0A3eE4E71c8681216E0C17231B & 4 & 0.6976665645274818 \\
			& 1PY5y459ErUiexiwhpWeCLRcBKfqtxagQk & 2 & 
			0.00209249 \\ \hline

		\end{tabular}
		\label{trs}
	\end{center}
\end{table*}

In total 151 ethereium transactions and 110 Bitcoin transactions submitted to the fake giveaways addresses.
It can be seen that the following domain names have used same eth/bitcoin addresses:
\begin{itemize}
	\item teslaget.org and ether2x.net
	\item muskelon.io and ms2x.one
	\item fundtesla.us and 2xtesla.us
\end{itemize}

It has been observed that 2xbitcoin.org has changed its eth address during the giveaway fraud. 

For Ethereum transactions, etherscan has translated the sender's address of several transactions to cryptocurrency exchange platforms' names. The number of identified transactions of the cryptocurrency exchange platform is listed as follows:
\begin{itemize}
	\item Coinbase - 41
	\item Binance - 16
	\item Kraken - 2
	\item LUNO - 2
	\item FTX exchange - 2 
	\item Crypto.com - 1 
	\item OKEx - 1 
	\item CEX.IO - 1 
\end{itemize}
 
Once the website/YouTube account is filtered, the scammers purged the stolen funds into crypto mixers to hide it and then withdraw the stolen funds. This can be observed by checking the destination addresses that the funds have been transferred.

\vspace{1mm}
\textbf{Blacklisted addresses}

Checking the bitcoin addresses on scam-alert.io and bitcoinabuse.com on 5/20/2022, 12 addresses have been correctly labeled as ``Giveaway Scam" while 3 addresses were classified as unknown indicating that there were no report for these addresses. 
For ethereum addresses, no resource was found to check the ethereum blacklisted addresses. 

Checking cryptoscamdb which is an open-source dataset tracking scam URLs and their associated addresses, none of the domain names and their corresponding addresses have been listed at the time of writing this paper.  

\subsection{Reporting}

The author has reported the scam videos to YouTube as soon as seeing the videos. This can be done by choosing the flag icon under the video, and then report the video using the correct classification. The author has selected ``Spam or misleading" $>$ ``Scams or fraud" category.
YouTube has removed the reported videos after around +8 hours on average (the time of removing for different videos were varied). It is expected that more than one viewer is reporting the scam videos specifically due to the large number of watchers. YouTube sent an email regarding this report:

\textit{"Hello,
Thank you for reporting videos you find inappropriate. The video that you reported to us on May 15, 2022 has been removed or restricted from YouTube.
Please note that Creators have the right to appeal YouTube's decision any time after their content is removed or restricted. You are able to check the status of the content you flagged at any time."
"}

Checking official authorities investigating the cryptocurrency frauds, the following sources have been found:
\begin{itemize}
	\item Federal Trade Commission (FTC) report fraud section, https://reportfraud.ftc.gov/
	\item U.S. Securities and Exchange Commission (SEC), https://www.sec.gov/tcr
\end{itemize}

The author has reported the scam websites/scheme to FTC and SEC using the above links. However, no reply has been received regarding these reports.

%
%

%
%
%
%
%
%
%
%
%

\section{Countermeasures}

In this section, the countermeasures that can be considered to protect users from cryptocurrency scams are discussed. 

\vspace{1mm}
\textbf{Educating users}

Arguably the most important factor in preventing such a simple scam scheme is to educate users regarding cyber scams. It seems many cryptocurrency owners are not well-educated regarding cyber-attacks and scams. Despite the wide cryptocurrency investment from diverse investors, there is a lack of sufficient education regarding the cybersecurity and cyber scams in this area causing a huge loss specifically for micro investors. 

On the other hand, although detecting a giveaway scam is common sense, still many users fail to detect such a simple fraud. This failing can be rooted in an incorrect mindset/culture developed in cryptocurrency communities that it is easy to become rich by investing in the crypto market. Therefore, the education should not be only limited to cybersecurity concepts but also cover basic economic rules behind the crypto market such that crypto owners can invest mindfully.

\vspace{1mm}
\textbf{Blacklist ETH addresses}

Surprisingly, at the time of writing this paper, the author has not found any intelligent source dedicated to reporting the fraudulent Ethereum addresses. 
For bitcoin, two active websites have been found which are scam-alert.io and bitcoinabuse.com. On these websites, users can simply report a fraudulent bitcoin address, and the website keeps a database of flagged fraudulent addresses. Users can inquire about bitcoin addresses on these websites. 
Having a database of scam Ethereum addresses can also effectively protect users by helping them to identify fraud addresses.

On the other hand, as many users are using cryptocurrency exchange platforms such as Coinbase, and Binance, it is expected that these companies actively check the reported scam addresses and alert users before committing the suspicious transactions. 

\vspace{1mm}
\textbf{YouTube removing process}

According to the Google transparency report, YouTube is actively removing Channels violating its policy~\cite{transp}. From Oct-2021 to Dec-2021, YouTube has removed 3,850,275 channels. It can be seen that more than \%90 of the detection of violating videos has been done by Automated Flagging, and then the rest is from Users, Individual Trusted Flaggers, NGOs, and Government agencies. However, following items can be considered to reduce the impact of live stream scam. 
\begin{itemize}
	\item YouTube has a verified badge for verified channels. It can also add ``Not-verified" badge for channels choosing the same name as verified channels such as ``Tesla". 
	
	\item YouTube can consider some incentives to motivate users in reporting contents violating its guideline. Currently, as there is no incentivization, many users are not participating in the reporting process.
	
	\item The time between flagging a scam video and removing it from YouTube is considerably long causing the scam to be successful. From the author's perspective detecting a cryptocurrency giveaway live stream scam is an easy task, and removing it should not take hours. 
	
	\item As YouTube is equipped with an Automated Flagging system to detect inappropriate videos, this tool can be developed to detect scam giveaway videos as well. 
	Considering the development of deep fake videos, it is expected that more advanced giveaway scams can be conducted in the near future with a fake video of a famous person requesting donations for a giveaway in the video.

\end{itemize}

\vspace{1mm}
\textbf{Regulations}

The lack of strict regulations in the area of cryptocurrency is another fact leading to such scam schemes. Regulations can put more pressure on big tech companies to take more responsibility for detecting and preventing scam schemes. 
Moreover, the regulations can enforce cryptocurrency exchange platforms to check the validity of cryptocurrency addresses before committing the transaction.

Recently Securities and Exchange Commission (SEC) has announced new crypto regulation initiatives to expand investor protections in the crypto market~\cite{sec}. SEC plans to register and regulate crypto exchanges.

\section{Conclusion}

This paper has analyzed the simple and effective cryptocurrency giveaway scam based on YouTube live stream. 
The analysis shows that using this technique, scammers are stealing a large amount of money in a short time.
The paper has also discussed the feasible countermeasures which can be applied to protect users from live stream cryptocurrency giveaway scams in the future.

\appendices 

\Urlmuskip=0mu plus 1mu\relax
\bibliographystyle{IEEETran}
\bibliography{pppp}

\begin{thebibliography}{10}
\providecommand{\url}[1]{#1}
\csname url@samestyle\endcsname
\providecommand{\newblock}{\relax}
\providecommand{\bibinfo}[2]{#2}
\providecommand{\BIBentrySTDinterwordspacing}{\spaceskip=0pt\relax}
\providecommand{\BIBentryALTinterwordstretchfactor}{4}
\providecommand{\BIBentryALTinterwordspacing}{\spaceskip=\fontdimen2\font plus
\BIBentryALTinterwordstretchfactor\fontdimen3\font minus
  \fontdimen4\font\relax}
\providecommand{\BIBforeignlanguage}[2]{{%
\expandafter\ifx\csname l@#1\endcsname\relax
\typeout{** WARNING: IEEEtran.bst: No hyphenation pattern has been}%
\typeout{** loaded for the language `#1'. Using the pattern for}%
\typeout{** the default language instead.}%
\else
\language=\csname l@#1\endcsname
\fi
#2}}
\providecommand{\BIBdecl}{\relax}
\BIBdecl

\bibitem{group}
``Scammers made 1.6 million dollars in yet another fake crypto giveaway,''
  \url{https://www.group-ib.com/media/crypto-trap/}, accessed: 2022-05-20.

\bibitem{tenable2}
``Elon musk and snl: Scammers steal over 10 million in fake bitcoin, ethereum
  and dogecoin crypto giveaways,''
  \url{https://www.tenable.com/blog/elon-musk-snl-scammers-steal-over-10-million-in-bitcoin-ethereum-dogecoin-crypto-fake-giveaways},
  accessed: 2022-05-18.

\bibitem{tenable}
``Fake bitcoin, ethereum, dogecoin, cardano, ripple and shiba inu giveaways
  proliferate on youtube live,''
  \url{https://www.tenable.com/blog/fake-bitcoin-ethereum-dogecoin-cardano-ripple-and-shiba-inu-giveaways-proliferate-on-youtube},
  accessed: 2022-05-18.

\bibitem{wozni}
``Apple co-founder steve wozniak can’t sue youtube over bitcoin scam,''
  \url{https://www.bloomberg.com/news/articles/2021-06-03/apple-co-founder-wozniak-can-t-sue-youtube-over-bitcoin-scam},
  accessed: 2022-05-20.

\bibitem{ripple}
``Ripple files lawsuit against youtube: "enough is enough",''
  \url{https://cointelegraph.com/news/ripple-files-lawsuit-against-youtube-enough-is-enough},
  accessed: 2022-05-20.

\bibitem{ripple22}
``Ripple ends youtube lawsuit over xrp giveaway scams, says ceo,''
  \url{https://cointelegraph.com/news/ripple-ends-youtube-lawsuit-over-xrp-giveaway-scams-says-ceo},
  accessed: 2022-05-20.

\bibitem{goog}
``Phishing campaign targets youtube creators with cookie theft malware,''
  \url{https://blog.google/threat-analysis-group/phishing-campaign-targets-youtube-creators-cookie-theft-malware/},
  accessed: 2022-05-20.

\bibitem{interv}
``The bitcoin word | live with cathie wood, jack dorsey, and elon musk,''
  \url{https://www.youtube.com/watch?v=Zwx_7XAJ3p0}, accessed: 2022-05-18.

\bibitem{interv2}
``Third row tesla podcast – episode 7 - elon musk's story - director's cut,''
  \url{https://www.youtube.com/watch?v=J9oEc0wCQDE}, accessed: 2022-05-18.

\bibitem{interv3}
``he best business show with anthony pompliano: Exclusive michael saylor
  interview,'' \url{https://www.youtube.com/watch?v=c3E91-RGjQE}, accessed:
  2022-05-18.

\bibitem{widget}
``Bitcoin widgets / crypto price widgets,'' \url{https://coinlib.io/widgets},
  accessed: 2022-05-20.

\bibitem{transp}
``Youtube community guidelines enforcement,''
  \url{https://transparencyreport.google.com/youtube-policy/removals?hl=en},
  accessed: 2022-05-18.

\bibitem{sec}
``The sec announced new crypto regulation initiatives this week. here’s what
  investors should know,''
  \url{https://time.com/nextadvisor/investing/cryptocurrency/sec-new-crypto-regulation-gensler/},
  accessed: 2022-05-18.

\end{thebibliography}

\end{document}